\documentstyle[12pt,graphics]{article}
\begin{document}

\title{New colour-transformations for the $Sloan$ photometry and revised 
metallicity calibration and  equations for photometric parallax 
estimation}
\author{S.~Karaali, 
 S.~Bilir, 
 and S.~Tun\c{c}el \and
 } 
\date{}
\maketitle

{\center
Istanbul University, Science Faculty, 
Department of Astronomy and Space Sciences, 34119 University -
Istanbul, Turkey \\karsa@istanbul.edu.tr\\
                  sbilir@istanbul.edu.tr\\
		tuncelsabiha@hotmail.com\\[3mm]
}

\begin{abstract}
We evaluated new colour-transformations for the $Sloan$ photometry by 224 
standards and used them to revise both the equations for photometric 
parallax estimation and metallicity calibration cited by Karaali et 
al. (2003). This process improves the metallicity and absolute 
magnitude estimations by $[Fe/H]\leq0.3$ dex and $M^{H}_{g^{'}}\geq0.1$ 
mag respectively. There is a high correlation for metallicities 
and absolute magnitudes derived for two systems, $UBV$ and $Sloan$, 
by means of the revised calibrations.
\end{abstract}


\bigskip

\section{Introduction}

In a recent paper, Karaali et al. (2003) presented a new procedure 
for the photometric parallax estimation and extended it to the $Sloan$ 
photometry. Also, they  marked the advantage of the new procedure 
relative to the ones already used. Especially, they noticed that this 
procedure would be more appropriate for stars of a population with a 
large metallicity range for which a single colour-magnitude diagram is 
used in the literature (cf. Chen et al. 2001, and Siegel et al. 2002). 

However, unlike the colour-transformations derived for broad band photometric 
systems (see, e.g., Buser 1978), the transformations derived for the 
$Sloan$ system (Fukugita et al. 1996) are functions of a single colour,  
i.e. the colour $(u^{'}-g^{'})$ is a function of only $(U-B)$, and 
$(g^{'}-r^{'})$ is another function of only $(B-V)$. This was the case about 
a decade ago (cf. Fukugita et al. 1996) and it is still the same for the 
very recent years (Smith et al. 2002). We thought to evaluate new 
colour-transformations for $(u^{'}-g^{'})$ and $(g^{'}-r^{'})$ which would 
cover both $(U-B)$ and $(B-V)$, by using the data of standards already used 
in the literature. This is the main scope of this work. We will show in the 
following sections that such an approach provides more precise absolute 
magnitudes (and metallicities) than those in the paper of Karaali et al. 
(2003).

Also, we wish to emphasize that the new photometric parallax and $[Fe/H]$ 
equations are only one example of the improvement in science that could 
result from these new transformations.

In Section 2 we present the data used for calibration and the new 
colour-transformations, and in Section 3 the new metallicity calibration is 
given. New equations for photometric parallaxes are given in Section 4 and 
in Section 5 a short conclusion is provided.

\section{New Colour-Transformations for the $Sloan$ Photometry}

We used the $UBV$ data of Landolt (1992), for 251 stars, whereas 
$u^{'}g^{'}r^{'}$ data for the same stars are provided from the WEB page of 
CASU INT Wide Field Survey\footnote{http://www.ast.cam.ac.uk/ $\sim$ wfcsur/index.php}.
Fig. 1 and Fig. 2 which plot $(g^{'}-r^{'})$ versus $(B-V)$ 
and $(u^{'}-g^{'})$ versus $(U-B)$ show that there are some stars at extreme 
locations in these diagrams. These stars either have $(B-V)$ colour indices 
less than 0.3 mag or larger than 1.1 mag where the calibration of Karaali et 
al. (2003) does not hold, or their standard errors for the colour indices 
cited above are relatively large. The total number of stars which lie at 
extreme locations in Figs. 1 and 2 is 27. We excluded them from the sample 
and we marked the remaining 224 points with agreeable colours into Fig. 3 
and Fig. 4. It is interesting, the points in each figure do not lie on a 
straight line, but they form a band. This shows schematically that 
$(u^{'}-g^{'})$ and $(g^{'}-r^{'})$ are functions of both $(U-B)$ and 
$(B-V)$.

\begin{figure}
\resizebox{7cm}{7cm}{\includegraphics*{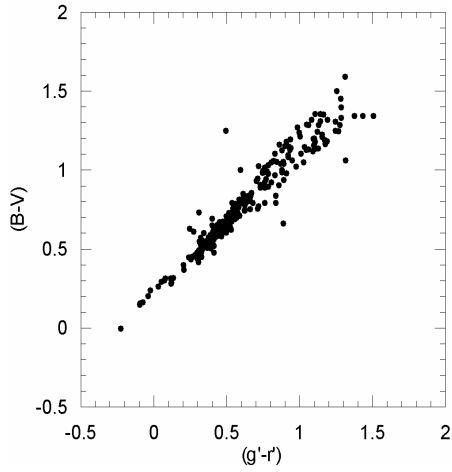}}  
\caption{$(g^{'}-r^{'})$ versus $(B-V)$ for the original sample of 
stars. There are some stars at extreme locations}
\end{figure}

\begin{figure}
\resizebox{7cm}{7cm}{\includegraphics*{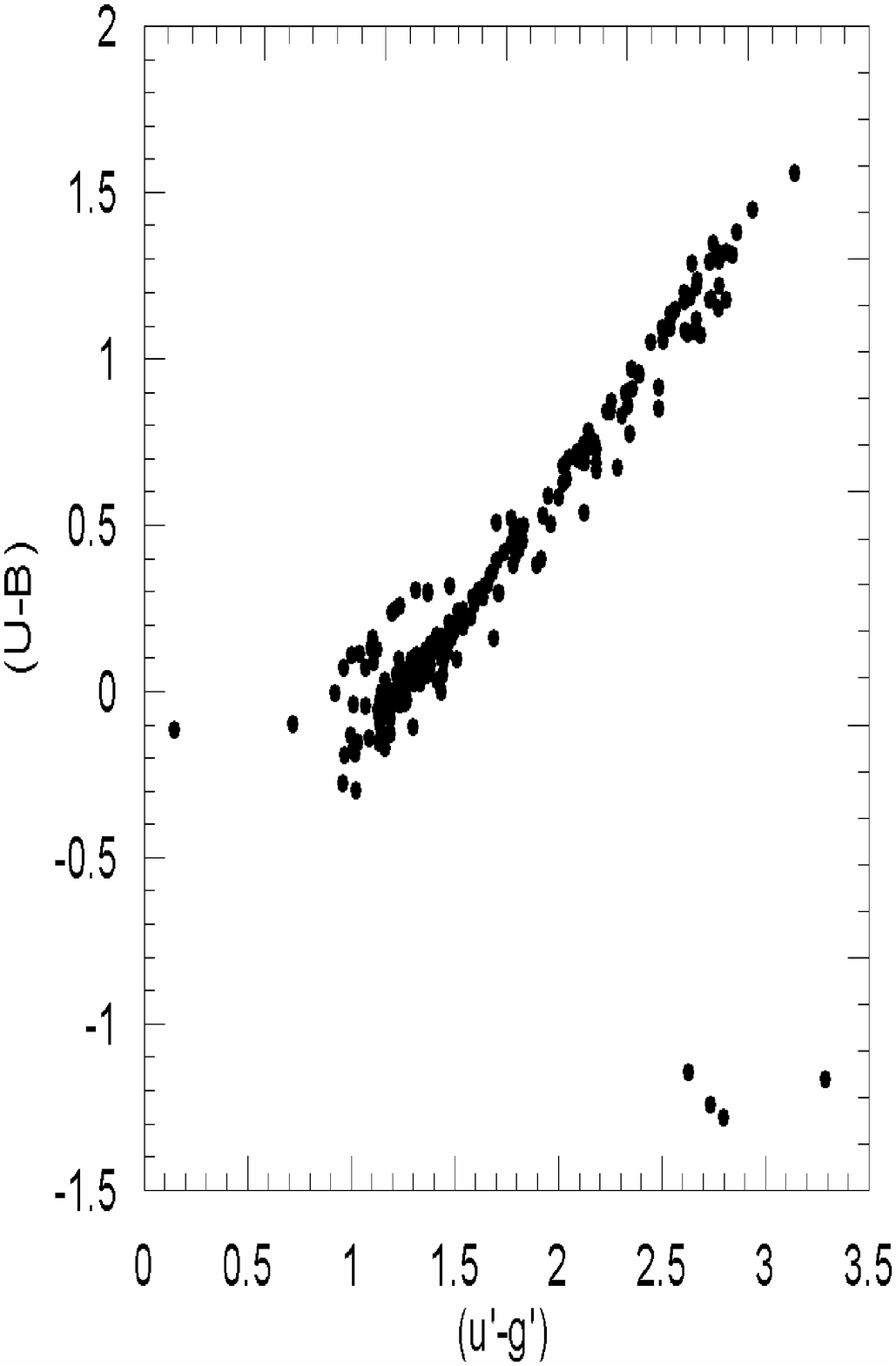}}  
\caption{$(u^{'}-g^{'})$ versus $(U-B)$ for the original sample of 
stars. There are some stars at extreme locations}
\end{figure}

\begin{figure}
\resizebox{7cm}{7cm}{\includegraphics*{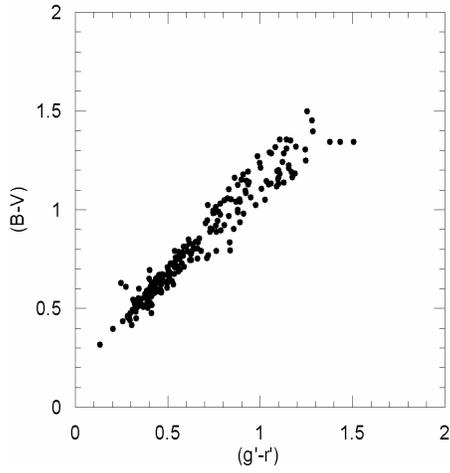}}  
\caption{$(g^{'}-r^{'})$ versus $(B-V)$ for 224 stars. 27 stars at 
extreme locations in Fig.1 are excluded from the sample}
\end{figure}

\begin{figure}
\resizebox{7cm}{7cm}{\includegraphics*{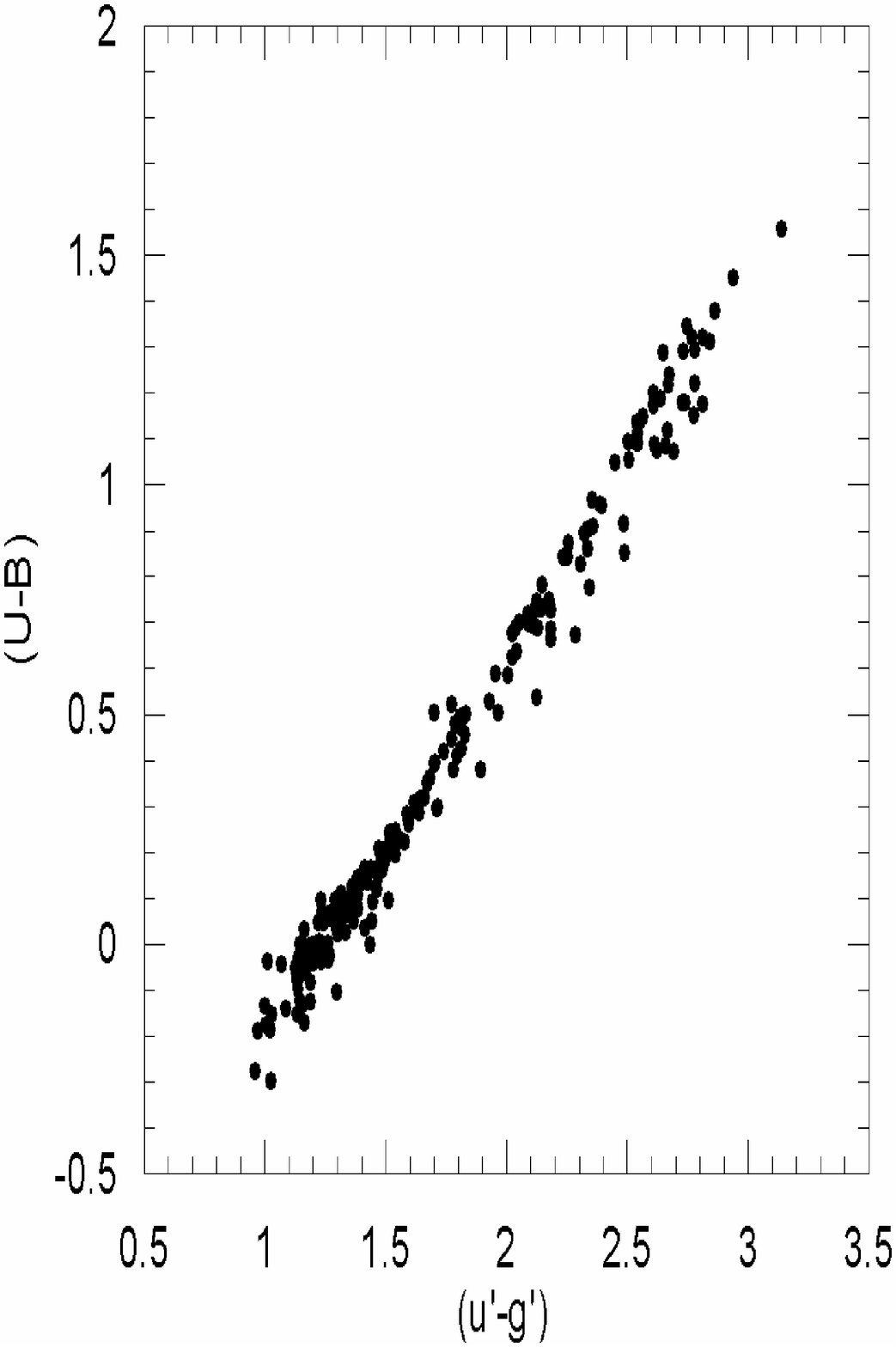}}  
\caption{$(u^{'}-g^{'})$ versus $(U-B)$ for the same sample in Fig. 3}
\end{figure}

We adopted the colour-transformations as follows and evaluated the 
coefficients by means of least square method:

\begin{eqnarray}
g^{'}-r^{'}=a(B-V) + b(U-B) + c\nonumber\\
u^{'}-g^{'}=d(U-B) + e(U-B) + f
\end{eqnarray}
Thus, we obtained the new colour-transformations for the $Sloan$ photometry 
as in the following:
\begin{eqnarray}
g^{'}-r^{'}=1.023(B-V) + 0.016(U-B) - 0.187 \nonumber\\
u^{'}-g^{'}=0.779(U-B) + 0.755(B-V) + 0.801 
\end{eqnarray}
We recall the colour-transformations of Fukugita et al. (1996) which were 
used in the paper of Karaali et al. (2003) for comparison purpose:

\begin{eqnarray}
g^{'}-r^{'}=1.05(B-V) - 0.23\nonumber\\
u^{'}-g^{'}=1.38(U-B) + 1.14					
\end{eqnarray}
It seems that the main difference between two set of equations is for 
$(u^{'}-g^{'})$, which is sensitive to metallicity and therefore results 
different absolute magnitudes for the same $(g^{'}-r^{'})$ colour index. 

To illustrate the difference between the two colour-transformations, we 
applied each set of transformations to the $UBV$ measures and compared the 
predicted $(u^{'}-g^{'})$ and $(g^{'}-r^{'})$ colours to the actual measures 
from the WEB page of CASU INT Wide Field Survey (Fig. 5 and Fig. 6). The mean 
of the differences between the evaluated $(u^{'}-g^{'})$ and the original ones 
is $<\Delta(u^{'}-g^{'})>=0.000$ mag for the new colour-transformation and 
$<\Delta(u^{'}-g^{'})>=0.044$ mag for the one of Fukugita et al. (1996), 
and the corresponding standard deviations are $s=\pm0.017$ mag and 
$s=\pm0.104$ mag, respectively. 

To demonstrate the superiority of our new transformations, we illustrate their 
impact on particular application - the derivation of photometry - based 
metallicity estimates. Karaali et al. (2003) estimate abundances based on the 
$\delta_{0.4}$ measure. A star of $\delta_{0.4}=0.25$ in the new transformation 
would have $\delta_{0.4}=0.294$ in the old transformation. Using the metallicity 
calibration given in Karaali et al. (2003),   

\begin{eqnarray}
[Fe/H]=0.10-2.00\delta_{0.4}-12.64\delta^{2}_{0.4}+11.43\delta^{3}_{0.4}
\end{eqnarray}
one obtains the metallicities $[Fe/H]=-1.01$ and $[Fe/H]=-1.29$ dex, for 
$\delta_{0.4}=0.25$ and $\delta_{0.4}=0.294$ respectively.

Additionally, the standard deviation $s=\pm0.104$ mag is roughly 
six times larger than $s=\pm0.017$ mag. That is, the colour-transformations of 
Fukugita et al. (1996) causes a larger dispersion relative to the new ones. 
All these differences cited here affect also the evaluation of absolute 
magnitudes and hence the distance to a star.

The evaluated $(g^{'}-r^{'})$ colour indices by means of two sets of 
colour-transformations are not as different as the $(u^{'}-g^{'})$ ones. The 
mean of the differences between the evaluated $(g^{'}-r^{'})$ and the original 
ones is $<\Delta(g^{'}-r^{'})>=0.000$ mag for equation (2), and 
$<\Delta(g^{'}-r^{'})>=0.027$ mag for equation (3). Although there is a small 
difference between these means, the corresponding standard deviations are 
almost equal, i.e. $s=\pm0.0700$ mag and $s=\pm0.0701$ mag for the new and old 
colour-transformations, respectively. Such  differences in $(g^{'}-r^{'})$ do 
not affect neither the metallicity nor the absolute magnitude estimation 
considerably.   

\section{Revised Metallicity Calibration}

We used the procedure given in the paper of Karaali et al. (2003) for revision 
the metallicity calibration. The only difference is between the different 
colour-transformations used. Let us write equation (2) for two stars 
with the same $(B-V)$ (equivalently $g^{'}-r^{'}$), i.e. for a Hyades 
star (H) and for a star ($\star$) whose UV-excess is normalized:

\begin{eqnarray}
(u^{'}-g^{'})_{H}=0.779(U-B)_{H} + 0.755(B-V) + 0.801\nonumber\\
(u^{'}-g^{'})_{\star}=0.779(U-B)_{*} + 0.755(B-V) + 0.801
\end{eqnarray}
Then, the UV-excess for the star in question, relative to the Hyades 
star is

\begin{eqnarray}
(u^{'}-g^{'})_{H}-(u^{'}-g^{'})_{\star}=0.779[(U-B)_{H}-(U-B)_{\star}]
\end{eqnarray}
or, in standard notation, 

\begin{eqnarray}
\delta(u^{'}-g^{'})=0.779\delta(U-B)
\end{eqnarray}
The $(U-B)$ colour index of a Hyades star with $(B-V)=0.6$ mag is $(U-B)_{H}=
0.13$ mag (Sandage 1969), and equation (2) transforms them to 
$(g^{'}-r^{'})=0.43$ mag. If we apply equation (7) to a star with $(B-V)=0.6$ 
mag, we obtain

\begin{eqnarray}
\delta(u^{'}-g^{'})_{0.43}=0.779\delta(U-B)_{0.6}		
\end{eqnarray}
for the relation between the normalized UV-excesses in $UBV$ and the 
$Sloan$ systems. From this equation we obtain 

\begin{eqnarray}
\delta(U-B)_{0.6}=1.284\delta(u^{'}-g^{'})_{0.43}  
\end{eqnarray}
which yields a revised metallicity calibration for the $Sloan$ 
photometry by its substitution in 
\begin{eqnarray}
[Fe/H]=0.10-2.76\delta_{0.6}-24.04\delta^{2}_{0.6}+30.00\delta^{3}_{0.6}	
\end{eqnarray}
which covers a large range of metallicity, i.e. $-2.75\leq[Fe/H]\leq0.2$ 
dex (Karaali et al., 2003). Hence, the revised metallicity calibration for 
the $Sloan$ photometry is obtained as follows:

\begin{eqnarray}
[Fe/H]=0.10-3.54\delta_{0.43}-39.63\delta^{2}_{0.43}+63.51\delta^{3}_{0.43}
\end{eqnarray}
We tested this calibration by comparison the metallicities for 155 stars, 
whose metal-abundances are in the range $-3.0\leq[Fe/H]\leq0.2$ dex, 
estimated by equations (10) and (11). The correlation is rather high, i.e. 
$[Fe/H]_{UBV}=0.9998[Fe/H]_{SLOAN}-0.006$, confirming the procedure used for 
derivation of the revised metallicity calibration for the $Sloan$ photometry.     

\begin{figure}
\resizebox{7cm}{7cm}{\includegraphics*{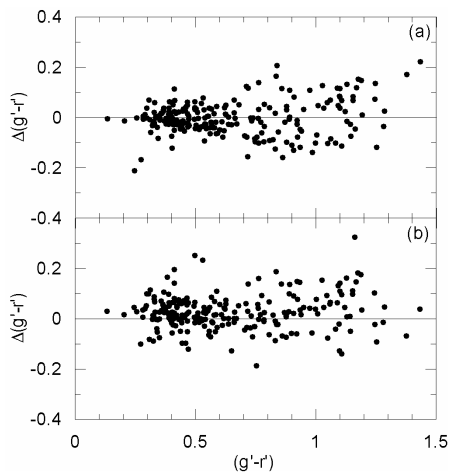}}  
\caption{Comparison of the predicted ($g^{'}-r^{'}$) colours with the actual 
measures. (a) for the new colour-transformations, and (b) for the 
colour-transformations of Fukugita et al. (1996)}
\end{figure}

\begin{figure}
\resizebox{7cm}{7cm}{\includegraphics*{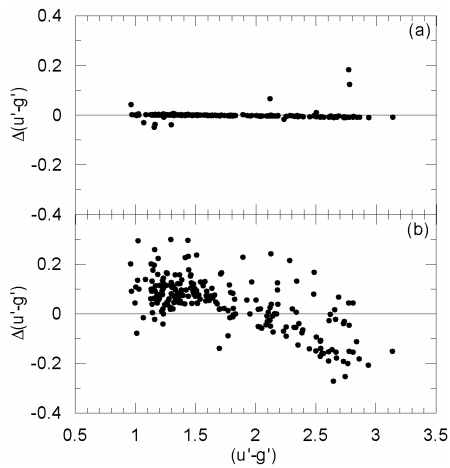}}  
\caption{Comparison of the predicted ($u^{'}-g^{'}$) colours with the actual 
measures. (a) for the new colour-transformations, and (b) for the 
colour-transformations of Fukugita et al. (1996)}
\end{figure}

\section{Revised Equations for the $Sloan$ Photometry for Photometric 
Parallax Estimation}

\subsection{The procedure}
We recall the procedure used in the paper of Karaali et al. (2003) for 
photometric parallax estimation for the $UBV$ photometry as follows:
They separated the stars into eight $(B-V)$ colour-index intervals, 
(0.3-0.4], (0.4-0.5], (0.5-0.6], (0.6-0.7], (0.7-0.8], (0.8-0.9], 
(0.9-1.0], and (1.0-1.1], and used the normalized equation 

\begin{eqnarray}
M^{H}_{V}(nor)=-2.1328(B-V)^{2}+8.6803(B-V)+0.305
\end{eqnarray}
for the fiducial main-sequence of Hyades as a standard main-sequence 
and derived the metallicity dependent offset 

\begin{eqnarray}
\Delta M^{H}_{V}(nor)=b_{3}\delta^{3}_{0.6}+b_{2}\delta^{2}_{0.6}
+b_{1}\delta_{0.6}+b_{0}
\end{eqnarray}
for each $(B-V)$ interval (we used the symbols in equation (13) the same 
as in the mentioned paper for consistency). The coefficients which were 
evaluated by 1236 stars are transfered here into Table 1. Thus, the 
$M_{V}(\star)$ absolute magnitude of a star with given $(B-V)$ colour 
index cited above can be estimated via 
$\Delta M^{H}_{V}(nor)=M_{V}(\star)-M^{H}_{V}(nor)$, provided that its 
$(U-B)$ colour index is available for $\delta_{0.6}$ normalized 
UV-excess evaluation.     

\begin{table}
\center
\caption{Numerical values for the coefficients of equation (13) as a 
function of $(B-V)$ colour index, transfered from Table 3 of 
Karaali et al. (2003)}
\begin{tabular}{crrrr}
\hline
$(B-V)_{o}$ &\multicolumn{1} {c} {$b_{3}$}&\multicolumn{1} {c} 
{$b_{2}$}&\multicolumn{1} {c} {$b_{1}$} &\multicolumn{1} {c} 
{$b_{0}$} \\
\hline
(0.3-0.4] &   -32.1800 &    15.9370 &     1.7350 &    -0.0177 \\
(0.4-0.5] &   -15.3820 &     3.7188 &     4.4850 &     0.0022 \\
(0.5-0.6] &     3.9109 &    -4.8075 &     5.3847 &     0.0134 \\
(0.6-0.7] &   -11.1700 &    -0.3015 &     5.0281 &     0.0153 \\
(0.7-0.8] &     0.1049 &    -3.6157 &     4.6196 &    -0.0144 \\
(0.8-0.9] &   -22.5350 &     0.1109 &     3.4469 &    -0.0203 \\
(0.9-1.0] &   -24.9710 &     7.2916 &     2.0269 &     0.0051 \\
(1.0-1.1] &    -7.4029 &     4.2761 &     1.2638 &    -0.0047 \\
\hline
\end{tabular}  
\end{table}

\subsection{Photometric Parallaxes}

We need to transform equations (12) and (13) into the $Sloan$ data. 
First we start with equation (12). Unfortunately, it is not as 
easy as in the case of colour-transformations of Fukugita et al. (1996), 
for $(B-V)$ is not only the function of $(g^{'}-r^{'})$ but also 
$(u^{'}-g^{'})$. However, the inverse colour-transformations in (2) can 
be obtained by mathematical calculations. The one for $(B-V)$ that will 
be used in equation (12) is as follows:

\begin{eqnarray}
(B-V)=0.992(g^{'}-r^{'})-0.0199(u^{'}-g^{'})+0.202
\end{eqnarray}		
If we replace the equivalence of $(B-V)$ in equation (14) into equation 
(12) and simplify it, we find the normalized colour-magnitude equation 
for the Hyades main-sequence as in the following:

\begin{eqnarray}		 				
M^{H}_{g^{'}}(nor)=-2.0987(g^{'}-r^{'})^{2}-0.0008(u^{'}-g^{'})^{2}
+0.0842(g^{'}-r^{'})(u^{'}-g^{'}) \nonumber \\
+7.7557(g^{'}-r^{'})-0.1556(u^{'}-g^{'})+1.9714
\end{eqnarray}
The transformation of equation (13) from $UBV$ to the $Sloan$ photometry 
is simple. It can be done by replacing the equivalence of $\delta_{0.6}$ 
in equation (9) into equation (13). The result is as follows:

\begin{eqnarray}	
\Delta M^{H}_{g^{'}}(nor)=c_{3}\delta^{3}_{0.43}+c_{2}\delta^{2}_{0.43}
+c_{1}\delta_{0.43}+c_{0}       
\end{eqnarray}
where $c_{3}=(1.284)^{3}b_{3}$, $c_{2}=(1.284)^{2}b_{2}$, 
$c_{1}=(1.284)b_{1}$, and $c_{0}=b_{0}$. The numerical data for $c_{i}$ 
(i= 0, 1, 2, and 3) are given in Table 2. 

\begin{table}
\center
\caption{Numerical values for the coefficients of equation (16) as a 
function of $(g^{'}-r^{'})$ colour-index. The colour-index intervals 
correspond to the $(B-V)$ intervals in the first column of Table 1}
\begin{tabular}{crrrr}
\hline
$(g^{'}-r^{'})_{o}$ &\multicolumn{1}{c}{$c_{3}$}&\multicolumn{1} {c} 
{$c_{2}$} & \multicolumn{1} {c} {$c_{1}$}& \multicolumn{1} 
{c} {$c_{0}$}\\
\hline
(0.12-0.22] &   -32.1800 &    15.9370 &     1.7350 &    -0.0177 \\
(0.22-0.32] &   -15.3820 &     3.7188 &     4.4850 &     0.0022 \\
(0.32-0.43] &     3.9109 &    -4.8075 &     5.3847 &     0.0134 \\
(0.43-0.53] &   -11.1700 &    -0.3015 &     5.0281 &     0.0153 \\
(0.53-0.64] &     0.1049 &    -3.6157 &     4.6196 &    -0.0144 \\
(0.64-0.74] &   -22.5350 &     0.1109 &     3.4469 &    -0.0203 \\
(0.74-1.85] &   -24.9710 &     7.2916 &     2.0269 &     0.0051 \\
(0.85-0.95] &    -7.4029 &     4.2761 &     1.2638 &    -0.0047 \\
\hline
\end{tabular}  
\end{table}

As mentioned in Section 2, the mean of the differences between predicted 
$(u^{'}-g^{'})$ and the actual $(u^{'}-g^{'})$, from the WEB page of 
CASU INT Wide Field Survey, is $<\Delta(u^{'}-g^{'})>=0.000$ mag for the 
new colour-transformations and $<\Delta (u^{'}-g^{'})>=0.044$ mag for the 
one of Fukugita et al. (1996). An excess of 0.044 mag in $\delta_{0.4}$ 
corresponds to an excess in absolute magnitude difference of $\sim0.1$ mag 
which can be confirmed as in Section 2. Actually, the evaluation of 
$\Delta M^{H}_{g^{'}}(nor)$ by means of the equation given in the paper of 
Karaali et al. (2003) (their equation 20) for $\delta_{0.4}=0.294$ mag 
gives an excess between 0.07 and 0.14 mag in absolute magnitude difference 
relative to $\delta_{0.4}=0.25$ mag, depending on the $(g^{'}-r^{'})$ colour 
index of the star. Hence, we can say that the revised equations for the 
photometric parallax estimation improve the absolute magnitudes by at least 
$\sim0.1$ mag. 

We compared the absolute magnitudes for the sample mentioned in Section 3, 
except three stars, derived by means of two systems, i.e. $UBV$ and $Sloan$. 
Stars excluded from the sample are relatively faint and have standard errors 
in $(u^{'}-g^{'})$ and $(g^{'}-r^{'})$ larger than 0.1 mag. The mentioned 
comparison is given in Fig. 7. Although we don't expect one to one 
correspondence due to different bands in two systems, the relation is rather 
high, again confirming our procedure, i.e. $M(V)=0.9972M(g^{'})-0.0460$.        

\begin{figure}
\resizebox{7cm}{7cm}{\includegraphics*{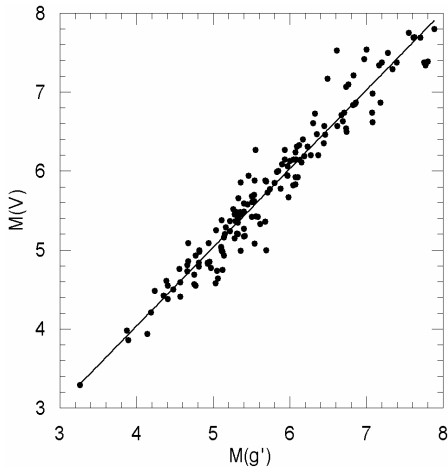}}  
\caption{Comparison of the absolute magnitudes $M(g^{'})$ and 
$M(V)$ derived by means of the procedure given in the text}
\end{figure}

\section{Conclusion}

Although the new procedure given in the paper of Karaali et al. (2003) 
provides significant improved photometric parallaxes with respect to the 
work of Laird, Carney, \& Latham (1988), it is compromised by the 
colour-transformations of Fukugita et al. (1996). Hence, we evaluated new 
colour-transformations for the $Sloan$ photometry by 224 standards and used 
them to revise both the equations for photometric parallax estimation and 
metallicity calibration. There is a high correlation for metallicities 
and absolute magnitudes derived for two systems, $UBV$ and $Sloan$, by 
means of the revised calibrations. This process improves the metallicity 
up to 0.3 dex (Section 2) and absolute magnitude at least 0.1 mag 
(Section 4). The improvements will provide better results in the works 
of Galactic structure related with the metallicity gradient and space 
density functions.

We should mention that the equations for $Sloan$ photometric parallax and 
$[Fe/H]$ would be significantly better if an analysis were made purely from 
$Sloan$ data. However, our aim is to confirm the improved of the conversion 
of $Sloan$ photometry to broad $UBV$ indices.

\end{document}